\documentclass[12pt]{article}
\input epsf.sty
\def\be{\begin{equation}} \def\ee{\end{equation}}
\def\bea{\begin{eqnarray}} \def\eea{\end{eqnarray}} \def\ba{\begin{array}}
\def\ea{\end{array}} \def\ben{\begin{enumerate}} \def\een{\end{enumerate}}

\newcommand{\eqn}[1]{(\ref{#1})}

\def\ov{\over}
\def\br{\nonumber\\}

\begin{document}
{}~
\hfill\vbox{\hbox{hep-th/0608144} \hbox{SINP/TNP/06-nn}}\break

\vskip .5cm
\centerline{\large \bf
Information geometry of $U(1)$ instantons on  self-dual manifolds
}
\centerline{\large \bf}

\centerline{  Rajesh R. Parwani}

\vspace*{.25cm}

\centerline{ \it Department of Physics, National University of Singapore} 
\centerline{ \it  Singapore}

\vspace*{.5cm}

\centerline{  Harvendra Singh}

\vspace*{.25cm}

\centerline{ \it Saha Institute of Nuclear Physics} 
\centerline{ \it  1/AF Bidhannagar, Kolkata 700064, India}

\vspace*{.5cm}

\vskip.5cm
\centerline{E-mail: parwani@nus.edu.sg, h.singh@saha.ac.in}

\vskip1cm
%DRAFT : \today \\
\centerline{\bf Abstract} \bigskip

  We use the information metric to investigate the moduli space of  a
$U(1)$ instanton on (anti)self-dual manifolds, 
  finding an $AdS$ geometry similar to that for the moduli space of a 
Yang-Mills instanton on flat space. We discuss 
 our results from the perspective of gauge/gravity duality.

\vfill \eject

\baselineskip=16.2pt

%%%%%%%%%%%%%%%%%%%%%%%%%%%%%%%%%%%%%%%%%%
%%%%%%%%%%%%%%%%%%%%%%%%%%%%%%%%%%%%%%%%%%%%%%
\section{Introduction}

The Anti-de Sitter/conformal field theory conjecture of Maldacena \cite{maldacena}
 relates gravity on a $d+1$ dimensional spacetime to a $d$ dimensional gauge theory, thus being a concrete realisation 
of the holographic idea \cite{holo} whereby the gauge theory lives on the boundary  of the bulk AdS manifold 
\cite{gubser,witten}. This duality has been extensively tested and has provided a  fruitful alternative understanding 
of various aspects of gauge theories, both qualitatively and quantitatively. The large  body of work is cited in some 
reviews, for example \cite{aharony}.  

One important aspect of the duality concerns the role of instantons \cite{instantons} in the boundary gauge theory. It 
was shown in Ref.\cite{dorey} that for a $SU(N)$ instanton, in the large $N$ limit, the moduli space equipped with the 
usual $L^2$-metric was equivalent to the bulk geometry. On the other hand,  the authors of Ref.\cite{blau} argued that 
the bulk spacetime  corresponding to a boundary $SU(N)$ theory could be identified  precisely with the one $SU(2)$-instanton 
moduli space equipped with an alternative metric. That is, in this latter proposal,  one does not need the limits taken 
in \cite{dorey} but one does need to adopt the information metric rather 
than the $L^2$-metric. This is an intriguing result, for it suggests that 
one may uncover, in a relatively simple way, the higher dimensional 
holographic bulk geometry corresponding to a boundary theory.

The information metric $G_{AB}$ is commonly used in probability theory and 
it associates a geometry to  
probability distributions \cite{fisher}, 
\be 
G_{AB} = \int d^d x \sqrt{g} P(x^\mu;\xi^A ) (\partial_A \ln P(x^\mu;\xi^A ))(\partial_B \ln P(x^\mu;\xi^A ) )  
\label{fisher}
\ee
where $P(x^\mu;\xi^A )$ is a probability density of $x^{\mu}$ while  
$ \xi^A, \; A=1,2,...,k $ are some continuous parameters. 
Hitchin \cite{hitchin} suggested the use of this metric for instanton  studies through the identification $P \to F^2$, 
with $F$ the instanton field strength. The advantage of the information  metric over the $L^2$-metric is its manifest 
gauge invariance and preservation of existent spacetime symmetries. 

For the usual $SU(2)$ Yang-Mills instanton in flat space, one has
\be
F^2 = {\rho^4 \over [ (x-x_0)^2 + \rho^2 ]^4} \label{f1}
\ee
 and so the information metric is \cite{grossier,blau}
 \be
 ds^2 =G_{AB} d \xi^A d \xi^B \sim { d\rho^2 + d (x^{\mu}_0)^2 \over 
\rho^2} \, ,
 \ee
 giving the moduli space an $AdS_5$ geometry. Deformations of the flat  base space of the gauge theory were studied in 
\cite{blau} giving rise to corrections to the $AdS_5$ geometry. Another  type of deformation of the base space that has 
been studied within this approach is the use of non-commutative space in  the limit of small non-commutativity 
\cite{parvizi}. There have also been studies of the information geometry of  supersymmetric theories \cite{britto}, 
lower dimensional field theories \cite{shi}, and at finite temperature \cite{rey}.  For earlier applications of information geometry in statistical physics, see \cite{stat}.
 
 In this paper we wish to study the information geometry of gauge 
instantons  living on a manifold which is not a small 
 deformation of flat space. We choose two of the simplest possibilities 
for our  investigation, a $U(1)$ instanton on a 
Eguchi-Hanson or Taub-NUT manifold. In the next two sections we review the 
pertinent aspects of Ref.\cite{eguchi} and then obtain 
the $U(1)$ instanton information geometry. In Section (4) we discuss the 
holographic aspects of our results and in the final section we discuss the implications.

\section{$U(1)$ instanton on the Eguchi-Hanson manifold}

The Euclidean gravity solutions with finite action and a self-dual 
 curvature are manifolds that are generally classified as gravitational 
instantons.  This is because the gravitational 
fields are 
localised in space, the metric becoming asymptotically locally flat at infinity. 
In the case of the Eguchi-Hanson (EH) manifold the metric is \cite{eguchi}
\be\label{eh}
ds_{EH}^2=( 1-{a^4\over r^4})^{-1}dr^2 +{r^2\over 4} 
(d\theta^2+ \sin^2\theta d\phi^2)+{r^2\over 4}(1-{a^4\over r^4}) 
(d\psi+ \cos\theta d\phi)^2
\ee
where $r^2=t^2+x^2+y^2+z^2$ and the coordinate ranges are 
\be\label{ranges}
a\le 
r\le\infty,~0\le\theta\le\pi,~0\le\phi\le 2\pi,~0\le\psi\le 2\pi .
\ee
Note that at $r=a$ the metric is completely regular with the topology of the manifold being $R^2\times S^2$. 
Let us briefly review some aspects of the EH solution. 
The relevant spin-connections \footnote{The vierbeins are
$e^0=( 1-{a^4\over r^4})^{-1/2}dr,~e^1= r \sigma_x,~ 
e^2=r\sigma_y,~e^3=r (1-{a^4\over r^4})^{1/2} \sigma_z$.} 
\bea
\omega^1_{~0}=\omega^2_{~3}=(1-{a^4\over 
r^4})^{1/2}\sigma_x,~~\omega^2_{~0}=\omega^3_{~1}=(1-{a^4\over 
r^4})^{1/2}\sigma_y,~~
\omega^3_{~0}=\omega^1_{~2}=(1+{a^4\over r^4})^{1/2}\sigma_z 
\eea
are anti-self-dual. The $\sigma_i$'s in the above expressions 
are the Cartan-Maurer 1-forms and obey the cyclic 
relation $d\sigma_x=2 \sigma_y \wedge \sigma_z$ \cite{eguchi}. 
It then follows that the curvature components are also anti-self-dual 
$R^1_{~0}=R^2_{~3},~R^2_{~0}=R^3_{~1},~R^3_{~0}=R^1_{~2}.$

As shown in \cite{eguchi}, a solution of the coupled Einstein-Maxwell field equations exists 
where the metric is taken to be the same as above and the 1-form gauge 
potential is locally
\be
A_{(1)}= {a^2\over r^2}\sigma_z= {a^2\over 2 r^2}(d\psi +  \cos\theta 
d\phi) \, .
\ee
Correspondingly one finds that the Maxwell field strength 
\be
F_{(2)}={2 a^2\over r^4}(e^3\wedge e^0 + e^1\wedge e^2)
\ee
is anti-self-dual.
Hence it is harmonic and the Euclidean energy-momentum tensor 
vanishes. Thus the Einstein equations retain their empty-space 
form, 
\be 
R_{\mu\nu} =0.
\ee
Using the above facts, we get for a $U(1)$ instanton  
\be
(F_{\mu\nu})^2={16 a^4\over r^8} \, . \label{u1}
\ee
Thus we observe $F\sim 1/r^4$ and so the asymptotic behaviour 
of this $U(1)$ instanton is  
similar to that of a Yang-Mills instanton on flat Euclidean space. 

 It is possible to express the $U(1)$ result in a form that is closer to 
 the $SU(2)$ result. To achieve this one makes  the following coordinate 
change \cite{eguchi}, 
\be\label{c1} 
\rho^4=r^4-a^4 \,
\ee
so that the EH metric \eqn{eh} in the new coordinates becomes
\be\label{eh2}
ds_{EH}^2=( 1+{a^4\over \rho^4})^{-{1\over2}}(d\rho^2 +\rho^2 \sigma_z^2)+ 
(1+{a^4\over \rho^4})^{1\over2} {\rho^2\over4}(d\theta^2+ \sin^2\theta 
d\phi^2)
\ee
with new coordinate ranges 
\be\label{ranges1}
0\le \rho\le\infty,~0\le\theta\le\pi,~0\le\phi\le 2\pi,~0\le\psi\le 2\pi .
\ee
The gauge potential and field strength are now given by
\bea
&&A_{(1)}= {a^2\over \sqrt{\rho^4+a^4}}\sigma_z \, ,\br
&&
F_{(2)}={2 a^2\over \rho^4+a^4}(e^3\wedge e^0 + e^1\wedge e^2).
\eea
Thus for a $U(1)$ instanton 
located away from the origin,  
\be
(F_{\mu\nu})^2={16 a^4\over (|x-x_0|^4+a^4)^2} \label{u2}
\ee
where we have taken $\rho^2=(x^i-x_0^i)^2 ~(i=1,2,3,4)$ everywhere 
including in the background \eqn{eh2}. 
(In the EH background the base space is $R^4$, so we have been able to 
introduce {\it four} new position parameters $x_0^i$.)
Equation \eqn{u2} is strikingly similar to the Yang-Mills gauge 
invariant field strength in \eqn{f1}, the only difference being 
the distribution of powers in the denominator. 

Now, since the scalar 
curvature is zero the on-shell Einstein 
action vanishes, so the entire gravitational contribution to the action 
 comes from the surface term which also vanishes asymptotically 
\cite{eguchi}. That is, the total 
gravitational action vanishes,
\be 
S[g]=0 \, .
\ee
However, the Maxwell action is finite, \footnote{
Note that $\sqrt{Det(g)}= {\rho^3\over 8} \sin \theta $, and  $ {1\over 
8\pi^2}\int \sin \theta d\theta d\phi d\psi=1$.} 
 \be
S[A]={1\ov 2.2!}\int d^4x \sqrt{g} (F_{\mu\nu})^2 = \pi^2 \, . \label{lagra}
\ee
Thus we may take 
\be
P(x^\mu; \xi^A) \equiv {1\over\pi^2} F^2
\ee
for our probability measure in computing the information metric, 
with  coordinates 
$\xi^A\equiv(\xi^0;\xi^\alpha)=(a; x_0^1,\cdots, x_0^4)$ 
parametrizing various directions in instanton moduli space. 
We find
\bea
&& \partial_{\xi^0}\log P= 4 \left({1\over \xi^0} -{2(\xi^0)^3\over 
|x-\xi|^4+(\xi^0)^4} \right) \br
&& \partial_{\xi^\alpha}\log P= {8(x-\xi)^2 (x-\xi)_\alpha\over 
|x-\xi|^4+(\xi^0)^4} \ .
\label{u3}
\eea
Using (\ref{fisher}) and  (\ref{u3}) we get the information 
metric as
\be 
 ds^2_{ADS_{EH}} \sim  { (d\xi^0)^2 + (d \xi^\alpha)^2 \over (\xi^0)^2}, 
\, \,\,\, \, \alpha=1,2,3,4. \label{ADS}
\ee
That is, the 
information geometry corresponding to the moduli space of a $U(1)$ instanton on a four-dimensional EH manifold is a five-dimensional 
Euclidean anti-de Sitter space, just as for the $SU(2)$ instanton on flat space.

\section{Instantons on Taub-NUT space}
We would like to study another example of a $U(1)$ instanton with 
self-dual curvature. 
The four-dimensional Taub-NUT Euclidean geometry is 
\cite{taubnut,eguchi1,eguchi}
\be\label{tn1}
ds_{NUT}^2={r+m\over 4(r-m)} dr^2 +
{(r^2-m^2)\over4}(d\theta^2+ \sin^2\theta d\phi^2)+
{m^2(r-m)\over r+m}(d\psi + \cos\theta d\phi)^2
\ee
with the coordinates having the ranges 
\be\label{tn2}
m\le r \le\infty,~0\le\theta\le\pi,~0\le\phi\le 2\pi,~0\le\psi\le 4\pi .
\ee
For large $r$ the topology of the manifold is $S^1\times R^3$, while near 
$r=m$, the apparent coordinate singularity which is removable, the 
topology becomes flat $R^4$ \cite{eguchi}.  The gauge potential in 
the background of \eqn{tn1} is
\be\label{tn3}
A_{(1)}= {r-m\over2( r+m)}(d\psi + \cos\theta d\phi)\ .
\ee
With that, we get
\be
(F_{\mu\nu})^2={16 \over (r+m)^4} 
\label{tn4}
\ee
which is normalised as
\be\label{tn5}
 {1\over 4}\int\limits_m^\infty dr (mr^2-m^3)\, (F_{\mu\nu})^2=1\ .
\ee
In the Taub-NUT case the radial coordinate $r$ is actually defined over a 
three-dimensional Euclidean base. Thus we take the radial 
coordinate $r^2=(x^i-x_0^i)^2$ in the Cartesian coordinates, where 
$x_0^i~(i=1,2,3)$ are three position 
variables. Like in the EH case we can define the information 
probability density as
\be\label{tn6}
P(x^\mu; \xi^A) \equiv {c\over\pi^2} F^2
\ee
where  
$\xi^A\equiv(\xi^0;\xi^\alpha)=(m;x_0^1,\cdots,x_0^3)$ 
parametrize various directions in moduli space and $c$ is an appropriate 
normalisation constant such that $\int d^4x \sqrt{g} P=1$. 
We now find
\bea\label{tn7}
&& \partial_{\xi^0} \log P= -{4\over |x-\xi|+\xi^0} \, , ~~~ 
\partial_{\xi^\alpha} \log P=  {4 \over 
(|x-\xi|+\xi^0)} {(x-\xi)_\alpha\over |x-\xi|} \, .
\eea
Using (\ref{fisher}) and  (\ref{tn7}) we get the information metric 
\be 
ds^2_{ADS_{NUT}} \sim  { (d\xi^0)^2 + (d \xi^\alpha)^2 \over (\xi^0)^2} 
\, , ~~~\alpha=1,2,3 \ . \label{ADS4}
\ee
which is a {\it four}-dimensional anti-de Sitter space.
Note the difference: the 
information geometry corresponding to the moduli space of a 
$U(1)$ instanton on a EH manifold  was a {\it five}-dimensional 
anti-de Sitter space \eqn{ADS}. Nevertheless the information metric 
\eqn{ADS4} has a 
constant negative curvature which is striking.

\section{Holography}
Here we wish  to investigate the holographic aspects of the information 
metrics obtained above. Let us consider first the EH space and define the function
\be\label{lap1}
\Phi:={u^4 \over (|\xi|^4 + u^4)^2 } \, 
\ee
where we have set $u\equiv \xi^0$ for simplicity. Then we obtain the 
following result for the Laplacian defined over the anti-de Sitter
space \eqn{ADS}
\bea\label{lap2}
\nabla_{ADS_5}\Phi&=& u^2[ \partial_u\partial_u -{3\over u} \partial_u 
+\nabla_{\xi^\alpha}] \Phi   \br 
&=& {u^4\over (\xi^4+u^4)^4} \left( 
(32u^4-48\xi^2u^2)(u^4-\xi^4)+32u^4\xi^4\right)
\eea
where  $\nabla_{\xi^\alpha}$ 
is the Laplacian over flat $\xi^\alpha$ four-space.
From this we deduce that
$F^2$ may not be interpreted simply as a 
boundary-to-bulk propagator as was the case with 
YM instantons in Ref.\cite{blau}.
However, consider the case of EH manifold in 
the asymptotic region where $\rho^2\gg a^2$. In 
this region the EH metric \eqn{eh2} becomes a locally flat (ALE) space
\be
ds^2\sim d\rho^2 + \rho^2 (\sigma_x^2+ \sigma_y^2 + \sigma_z^2) \, .
\ee
We can equivalently consider the small instanton-size  
limit  $a\to 0$ of \eqn{eh2}.
In the holographic $AdS$ coordinates, $\xi^A$, it would correspond to
$u\to 0$ (or 
$|(x-\xi)|^2\gg u^2$), \footnote{Given our $AdS$ metric 
\eqn{ADS},
$\xi^0\to 0$
would correspond to the UV region while $\xi^0\to\infty$ would define the
IR limit for the boundary theory. Note that $\xi^0=0$ is also the boundary 
of the $ADS$ space.}   so  we
get from \eqn{u2}
\be
F^2\sim {u^4\over |x-\xi|^8 } \, 
\ee
and in this approximation we get the bulk Laplacian equation
\bea\label{hol5}
\nabla_{AdS} (F^2)^p 
                 &=& 4p(4p-4) (F^2)^p + 8p(8p-2) {u^2\over \xi^2} (F^2)^p \, .
\eea
So, if we work up to the leading order ($u\to 0$) we can write \eqn{hol5} 
as  
\be\label{hol6}
\nabla_{ADS} (F^2)^p 
                 \approx 4p(4p-4)  (F^2)^p +{\cal O}({u^2\over |\xi|^2}) \ 
.
\ee
Also since 
\be
\lim_{u \to 0} F^2\sim\delta(x-\xi)\  ,   
\ee
this means that if we  concentrate near the AdS 
boundary, $u \to 0$, then $(F^2)^p$ does indeed act like a
boundary-to-bulk propagator 
for a massive scalar field. 
Specially for $p=1$ eq.\eqn{hol6} defines massless scalar field 
propagation in the bulk, just as in Ref.\cite{blau}. 

In the leading order,
we also recover the identity
\be\label{hol7}
{1\over 16} G^{AB} \partial_A \log F^2 \partial_B \log F^2=1+
{\cal O}({u^2\over \xi^2}) \ 
\ee
that was noted in Ref.\cite{blau}.

For Taub-NUT space \eqn{tn1} we shall concentrate near the 
region  $r=m$. One can define a
new coordinate patch \cite{eguchi}
$$d\tau={1\over 2} \sqrt{r+m\over r-m} dr\ .$$ 
Then the metric \eqn{tn1} becomes that of a flat $R^4$ space 
\be
ds^2\sim d\tau^2 + \tau^2 (\sigma_x^2+ \sigma_y^2 + \sigma_z^2) \, .
\ee
Expanding as
 $r=m+\epsilon$, with $\epsilon \approx {\tau^2\over 2m}$, we also obtain
from \eqn{tn4}
\be
F^2\sim{16 (2m)^4\over (\tau^2 + (2m)^2)^4 }\ .
\ee
The last two equations are the same as for YM instantons \eqn{f1}
in flat space where the information metric is $AdS_5$.
 So it is obvious that for the Taub-NUT instantons the information metric 
is $AdS_5$ and holographic 
when $r\sim m$,
but the bulk is $AdS_4$ when $r>m$.
That means, the information metric for the Taub-NUT $U(1)$
instantons in the $r\to m$ limit flows from  $AdS_4$ space to
the $AdS_5$. This should not appear
unusual because the Taub-NUT space has
Type-IIA string (D6-brane)
as well as M-theory Kaluza-Klein monopole interpretation depending
upon the size of the fibered $\psi$-direction in \eqn{tn1}.

\section{Discussion}
We have studied  two of the simplest examples of gauge theory on 
nontrivial space.  We found, quite remarkably, that for a $U(1)$ instanton 
on the EH manifold,  the information geometry of the moduli space is not 
flat but instead has  constant negative curvature just as for the bulk 
theories in the usual  gauge-gravity correspondence \cite{maldacena}. 
However our result does not have exactly the same 
bulk-boundary link that was found for the usual case in \cite{blau}. 
There are two ways to 
see this. Firstly, the boundary  of $AdS$ space is flat whereas our base 
theory (the supposedly boundary  theory) is a curved (anti)sef-dual 
manifold. 
Secondly, in 
our case we find 
that our $F^2$ may not be  interpreted simply as a 
bulk-to-boundary scalar propagator as was the case with 
YM instantons. However, in the leading order 
approximation
${\cal O}({u^2\over |\xi|^2})$ as $u\to 0$, 
 we do recover the holography of 
Ref.\cite{blau}. The same is true for the 
Taub-NUT case: There also when 
we work in the region $r\to m$, we get to the usual holographic  
interpretation.  
 
That is, near the flat regions of the curved manifolds we studied,  we do 
have a holographic interpretation of the bulk information geometry  
similar to Ref.\cite{gubser, witten, blau}. {\it The novel aspect of our 
result 
is that the gauge fields in our case are abelian in contrast to the 
Yang-Mills fields in the usual studies.}     

Our results lead us to the natural suggestion that the idea of 
Ref.\cite{blau}  may 
perhaps be extended: Any gauge theory on a manifold, 
 not just non-abelian ones on nearly flat  manifolds, is dual to a gravity 
theory defined by the information geometry of 
the instantons of the gauge theory.  Admittedly  we have no evidence for 
this, except for the special cases mentioned in  Section(1) which fall 
 within the Maldacena framework and the limiting cases mentioned above. 
The results of the previous section 
indicate that such an extended  duality would  
need to be more general than the boundary-bulk  mapping discussed in Refs.
 \cite{gubser, witten, blau}, reducing in some limit to the usual flat (or 
near flat) manifold form. 

Finally,  we remark that without the  Maxwell field in the EH background 
we  do not have an obvious suitable candidate for the probability density 
in (\ref{fisher}) because the pure  gravity sector, although representing 
a nontrivial manifold, gives zero scalar curvature. But in the pure gravity 
case perhaps it might be interesting to study some other  candidate, such 
as the density for the Euler characteristic ($\chi$). In this way one might 
be able to study the information geometry of the {\it gravitational} 
instantons.

\section*{Acknowledgements}
We thank the High Energy Theory Group at ICTP, Trieste (Italy) for 
hospitality during  the final stages of this work. 
H.S. is thankful to K.S. Narain for a useful discussion and to the National 
University of Singapore  for  hospitality when this work was initiated,
while R.P likewise thanks the Saha Institute for hospitality during 
the intermediate stages of this project.

\end{document}